\documentclass[a4paper, 11pts]{article}

\usepackage[margin=2cm]{geometry}
\usepackage{graphicx}
\usepackage{afterpage}
\usepackage{amsmath}
\usepackage{enumitem}
\usepackage{hyperref}
\usepackage{xcolor}
\usepackage{titlesec}
\usepackage{lipsum}
\usepackage[utf8]{inputenc}
\usepackage[T1]{fontenc}
\usepackage{geometry}
\usepackage{graphicx}
\usepackage{dblfloatfix}
\usepackage{authblk}
\usepackage{multicol}
\usepackage{aas_macros}
\usepackage{natbib}   
\usepackage{hyperref} 
\usepackage{authblk}  
\usepackage{txfonts}  
\usepackage{etoolbox}

\definecolor{esoBlue}{HTML}{005AA7}
\definecolor{esoLightBlue}{HTML}{58C0E0}

\titleformat{\section}
  {\normalfont\large\bfseries\color{esoBlue}}
  {\thesection}{1em}{}

\titleformat{\subsection}
  {\normalfont\bfseries\color{esoLightBlue}}
  {\thesubsection}{1em}{}

\makeatletter
\newcommand{\maketitlepage}{%
\begin{titlepage}
\centering
{\LARGE\bfseries Unlocking the dynamics of Young Stellar Objects: \\\vspace{0.25cm} \textit{Time-Domain Interferometry with six 4-m class telescopes}\vspace{1cm}}

{\large
\renewcommand\Authsep{, }
\renewcommand\Authand{ et }
\renewcommand\Authands{, et }
\@author
\par}

\vfill
{\large December 15, 2025}
\end{titlepage}
}
\makeatother

\author[1]{A. Soulain}
\author[1]{B. Lopez}
\author[1]{A. Matter}
\author[2]{F. Lykou}
\author[3]{P. Boley}
\author[3]{M. Scheuck}
\author[3]{R. van Boekel}
\author[4]{J.-C. Augereau}
\author[4]{M. leTessier}

\author[4]{J. Bouvier}
\author[1]{P. Berio}
\author[2]{P. \'Abrah\'am}
\author[13]{N. Anugu}
\author[4]{J.-P. Berger}
\author[1]{R. Burn}
\author[5]{W.-C Danchi}
\author[6]{W.J. de Wit}
\author[1]{F. Drewes}
\author[1]{V. Fleury}
\author[12]{V. Hocdé}
\author[7]{W. Jaffe}
\author[2, 4]{\'A K\'osp\'al}
\author[6,8]{E. Koumpia}
\author[4]{J.-B. Lebouquin}
\author[10]{J. S. Martin}
\author[1]{H. Meheut}
\author[1]{F. Millour}
\author[1]{N. Nardetto}
\author[11]{E. Pantin}
\author[4]{K. Perraut}
\author[1]{R. Petrov}
\author[7]{L.N.A van Haastere}
\author[2]{J. Varga}
\author[9]{G. Weigelt}
\author[10]{S. Wolf}

\affil[1]{Universit\'e C\^ote d’Azur, Observatoire de la C\^ote d’Azur, CNRS, Laboratoire Lagrange, Nice, France}
\affil[2]{Konkoly Observatory, HUN-REN Research Centre for Astronomy and Earth Sciences, Konkoly Thege Miklós út 15-17,
1121 Budapest, Hungary}

\affil[3]{Max Planck Institute for Astronomy, Königstuhl 17, D-69117 Heidelberg, Germany}
\affil[4]{Université Grenoble Alpes, CNRS, IPAG, 38000 Grenoble, 
France}
\affil[5]{NASA Goddard Space Flight Center, Greenbelt, MD 20771, USA}
\affil[6]{ESO, Alonso de Córdova 3107, Vitacura, Casilla, 19001, Santiago, Chile}

\affil[9]{Max-Planck-Institut für Radioastronomie, Auf dem Hügel 69, 53121, Bonn, Germany}

\affil[7]{Leiden Observatory, Leiden University, Einsteinweg 55, 2333 CC Leiden, The Netherlands}

\affil[8]{ALMA Observatory, Alonso de Córdova 3107, Vitacura, Santiago, Chile}

\affil[10]{Institut für Theoretische Physik und Astrophysik, Christian-Albrechts-Universität zu Kiel, Leibnizstraße 15, 24118 Kiel, Germany}

\affil[11]{Université Paris-Saclay, Université Paris Cité CEA, CNRS, AIM, 91191, Gif-sur-Yvette, France}

\affil[12]{Nicolaus Copernicus Astronomical Centre, Polish Academy of Sciences, Bartycka 18, 00-716 Warszawa, Poland}

\affil[13]{The CHARA Array of Georgia State University, Mount Wilson Observatory, Mount Wilson, CA 91023, USA}

\begin{document}

\maketitlepage


\begin{center}
    {\Large \bfseries Unlocking the Dynamics of Young Stellar Objects}\\[0.4cm]
{\large \bfseries \centering Abstract}\\[0.3cm]
\end{center}

\parbox{0.92\textwidth}{

The dynamics of the inner regions of young stellar objects (YSOs) is driven by a variety of physical phenomena, from magnetospheres and accretion to the dust sublimation rim and inner disk flows. These inner environments evolve on timescales of hours to days, exactly when bursts, dips, and rapid structural changes carry the most valuable information about star and planet formations, but remain hardly reachable with current facilities. A better reactive infrastructure with six or more telescopes, combined with alerts from large time-domain surveys (e.g., at the era of LSST/Rubin type facilities), and equipped with instruments spanning from the $V$-band to the thermal infrared ($N$), would provide the instantaneous uv-coverage and spectral diagnostics needed to unambiguously interpret and image these events as they happen. Such a world's first time-domain interferometric observatory would enable qualitatively new science: directly linking optical and infrared variability to spatially resolved changes in magnetospheric accretion, inner-disk geometry, and dust and gas dynamics in the innermost astronomical unit. Crucially, connecting these processes to outer-scale unresolved information from JWST, ALMA, and the ELT would yield a complete tomography of the planet-forming region.
}
\vspace{0.4cm}

\begin{multicols}{2}

\section{The dynamic inner au: Resolving the physics of star/planet formation in 4D}

\begin{figure*}[t!]
     \centering
     \includegraphics[width=0.98\textwidth]{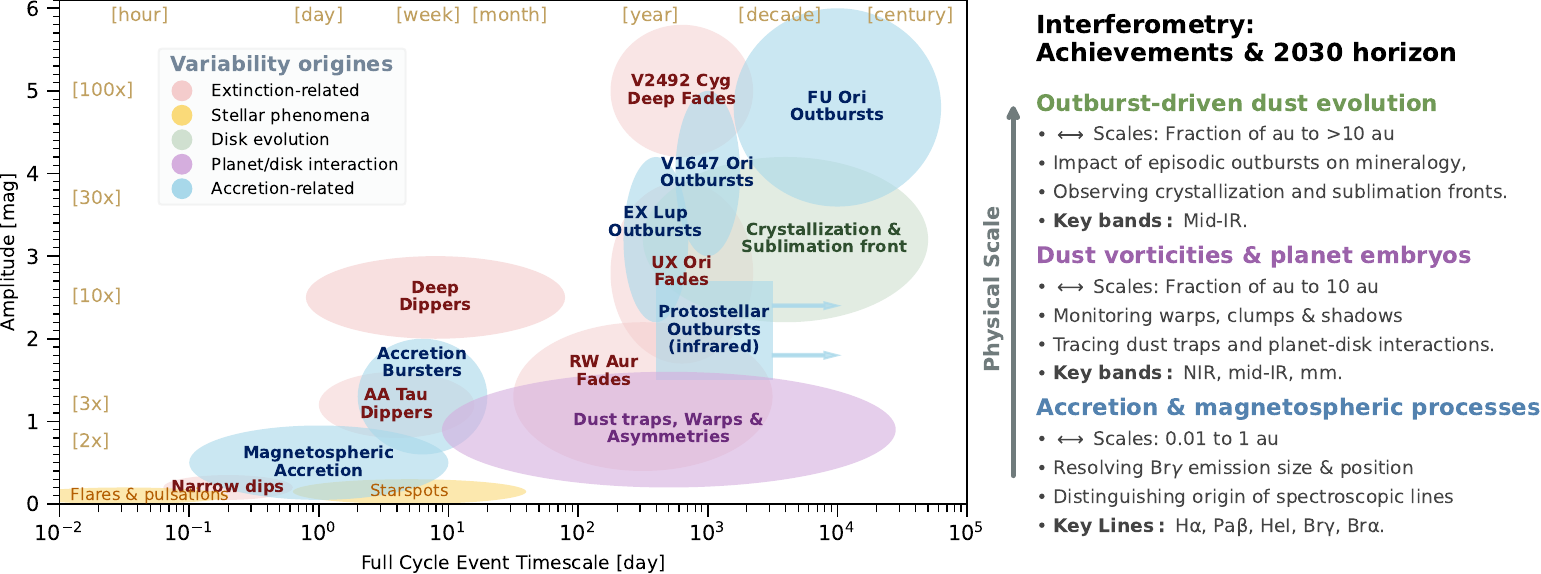}
     \vspace{-0.4cm}
     \caption{\small YSOs exhibit diverse, rapid, multi-scale variability modes that challenge the temporal resolution of existing arrays \citep[adapted from][]{Fischer2023}. Unveiling their origin requires capturing the inner geometry simultaneously with the observed variability.}
     \vspace{-10pt}
     \label{fig:variability}
 \end{figure*}

The innermost au scales of Young Stellar Objects (YSOs) is a complex interplay between magnetic fields, gas flows, and dust dynamics. While current optical interferometry has successfully resolved the static architecture of inner disks and magnetospheres, we possess limited spatially resolved information on the dynamic processes driving their evolution. Figure \ref{fig:variability} illustrates the vast dynamic range of YSO variability in both timescales and amplitudes. We structure our science case around four key regimes where resolving these temporal signatures spatially is critical.

\vspace{-0.3cm}
\subsection{Unraveling the accretion/ejection mechanism} The magnetosphere governs the early evolution of stars by controlling angular-momentum loss and channeling accretion \citep{Hartmann16}. However, the coupling between the star and the disk is highly unstable \citep{Pantolmos20, Zhu25}. We observe accretion bursts, rapid photometric dips, jets, winds and stochastic variability, but we cannot currently see the structural changes that cause them.
\begin{itemize}[label=\textendash, leftmargin=*, noitemsep, topsep=4pt]
    \item \textbf{The geometry of bursts:} Do accretion bursts result from large-scale disk instabilities or magnetic events?
    \item \textbf{Accretion in close binaries:} Resolving how binarity regulates accretion variability and shapes architecture is critical, as most stars form in multiple systems.
    \item \textbf{Magnetic funnels:} How do accretion funnels migrate and reconfigure on rotational timescales?
    \item \textbf{The ``black box'' of spectroscopy:} While it tells us \textit{how fast} gas is moving, it cannot tell us \textit{where} it is flowing.
\end{itemize}

\vspace{-0.5cm}
\subsection{The "Dipper" mystery: warps, clumps, or winds?} 

A large fraction of disk-bearing young stars exhibit day-to-week long quasi-periodic drops in optical brightness, suspected to be due to a dusty warp in the inner disk \citep[dipper phenomenon,][]{Roggero21, Nagel24}. However, without time-resolved imaging, we cannot confirm the origin of these dips. Is the extinction caused by a coherent geometric warp, a transient dust lift, or a dusty disk wind? Understanding Dippers is crucial as they likely represent the standard geometry of the star-disk interaction, yet their dynamic inner structure remains spatially unresolved during the dimming events.



\subsection{Transient signatures of planet formation} Beyond the inner rim, the planet-forming region is rich in transient phenomena that evolve on daily to monthly timescales. \begin{itemize}[label=\textendash, leftmargin=*, noitemsep, topsep=4pt] \item \textbf{Dynamics of disks:} Tracking the motion of vortices and dust traps triggered by the disk evolution or planet-disk interactions \citep{Varga21, Kuo24}. \item \textbf{Circumplanetary Disks (CPDs):} The formation of giant planets involves their own accretion disks. Detecting the moving variable thermal emission of a CPD against the stellar glare would provide a direct probe into the birth of gas giants \citep{Benisty21, Zhou25}. \item \textbf{Dust evolution and the 'Cosmic Furnace':} We observe sublimation cycles, re-condensation, and crystallization sites \cite[e.g., in EX Lupi outbursts,][]{Kospal23}. Spatially resolving these 'dust reactors' is crucial to understand how the ISM is turned into planetary building blocks (CAIs, chondrules), ultimately setting the chemical budget inherited by planets \citep{Kruijer20, Morbidelli24}. \end{itemize}

\vspace{-0.3cm}
\subsection{Extreme variability: The FUor/EXor eruptions}
Eruptive stars (FUors/EXors) represent the most violent manifestation of YSO dynamics which can occur for class I eruptions during a short time, with brightness increases up to 6 magnitudes \citep{Contreras24}, on timescales shorter than one year \citep{Laznevoi25}. These events induce profound structural changes that only reactive interferometry can resolve:
\begin{itemize}[noitemsep, topsep=0pt, leftmargin=*]
    \item \textbf{Magnetospheric Crushing:} High accretion rates ($>10^{-5} M_\odot/\text{yr}$) theoretically ``crush'' the magnetosphere, forcing a transition from funnel-flow to boundary-layer accretion \citep{Liu22}.
    \item \textbf{Thermal Restructuring:} The outburst launches a thermal wave, rapidly pushing the water snowline outward \citep{Cieza16} and triggering in-situ dust crystallization \citep{Abraham09}.
\end{itemize}
While surveys have tripled the known population \citep{Contreras24}, the triggering mechanism remains debated \citep{Nayakshin23}. We need to image these targets \textit{during} their brightness rise to distinguish between a propagating instability and a localized impact.

\subsection{The missing dimension: time} 
\vspace{-0.1cm}

The fundamental barrier to understanding these processes is not just spatial resolution or sensitivity, but temporal resolution. These environments evolve on timescales of hours to days -- exactly the domain where bursts, dips, and structural changes carry the most information. To understand the physics of star and planet formation, we must be able to image the geometry of gas and dust as it changes, linking the variable flux directly to the evolving structure.

\vspace{-0.3cm}
\subsection{Broader impact: A facility for the transient sky}
\vspace{-0.1cm}

While time-domain photometry (Gaia, ZTF, LSST) provides the alerting trigger and the period \citep{Hodgkin21}, only time-domain high angular resolution interferometry can capture the changing geometry. Only high-cadence monitoring at high spatial resolution can turn light curves into physical radii in the case of novae expansion \citep{Schaefer14} or disentangle static envelopes from dynamic mass-loss \citep[Cepheids,][]{Hocde25}. This versatility ensures that a $\ge6$ telescope array becomes a cornerstone of ESO's time-domain ecosystem, extending its legacy value far beyond YSOs.

\vspace{-0.3cm}
\section{Sensitivity is not the bottleneck}
\vspace{-0.2cm}

While the VLTI has achieved remarkable success in resolving the static architecture of bright objects, the transition to time-domain astronomy reveals fundamental structural limitations. The issue is no longer just seeing faint targets -- a frontier being pushed by GRAVITY+ \citep{GRAVITYplus2022} -- but rather observing them fast enough and often enough to disentangle between the different models hypothesis and constrain their physics.

\vspace{-0.3cm}
\subsection{The ``Time-Blurring'' barrier}
\vspace{-0.1cm}

Standard interferometric imaging relies on Earth-rotation synthesis to fill the $uv$-plane. However, for dynamic YSOs, this assumption of a static target breaks down. Integrating over a night or months effectively ``blurs'' these dynamical signatures. To freeze these events, we need to secure robust constraints on the object's geometry on timescales shorter than the variability itself. This requires \textbf{snapshot imaging capabilities} which is physically impossible with only 4 telescopes and movable configurations spread over several weeks.

\subsection{Operational rigidity \& sensitivity}
\vspace{-0.1cm}

Operationally, the 8-m Unit Telescopes (UTs) are too oversubscribed (1 week/month availability) to support monitoring campaigns or rapid response to alerts \citep{Manara21}. Conversely, the available 1.8-m Auxiliary Telescopes (ATs) or the CHARA array lack the sensitivity ($\sim12\%$ of close-in YSOs, Fig. \ref{fig:catalog}). We have to chose between sensitive telescopes that are unavailable and available telescopes that are not sensitive enough.

\vspace{-0.3cm}
\section{Golden era of Time-Domain astronomy}
\vspace{-0.2cm}

Looking toward the 2035+ horizon, the project is timely positioned to harvest the legacy of the Vera C. Rubin Observatory and the Roman Space Telescope. By then, these surveys will have cataloged the entire \textit{zoo} of YSO variability, delivering millions of light curves \citep{Ivezic19}.

\vspace{-0.3cm} 
\subsection{The discovery vs. characterization gap}
\vspace{-0.1cm}

However, such facilities are discovery engines, not a physical characterization engine. It identifies \textit{that} an event has occurred, but it cannot resolve its origin. A sudden brightening could be a magnetic reconnection event or a clump accretion; a dip could be a warped disk or a dust cloud. \textbf{To understand these events, we must spatially resolve them}. Capturing the instantaneous geometry of the accretion flow and the dynamical inner disk structure mandates a reactive array capable of immediate response to external alerts.

\vspace{-0.3cm} 
\subsection{The Keystone of the 2040 Landscape}
\vspace{-0.1cm}

In the 2040 ecosystem, this facility acts as the critical link between the thermal jets and cold disks monitored by SKA and the X-ray flares detected by Athena. By resolving the central core, it serves not merely as a follow-up machine, but as the indispensable spatially-resolved counterpart to the next generation of Great Observatories (ELT, HWO).

\vspace{-0.3cm}
\section{Six (or more) 4-m class telescopes: Agility, sensitivity, and instantaneous $uv$-coverage}
\vspace{-0.2cm}

To resolve the dynamical machinery of YSOs, we propose a paradigm shift: either (1) upgrading the VLTI auxiliary array with six or more 4-m class telescopes operating on baselines $>200$ m with visible-to-mid-IR instrumentation; or (2) developing a new interferometric network.

\begin{figure*}[t!]
     \centering
     \includegraphics[width=0.9\textwidth]{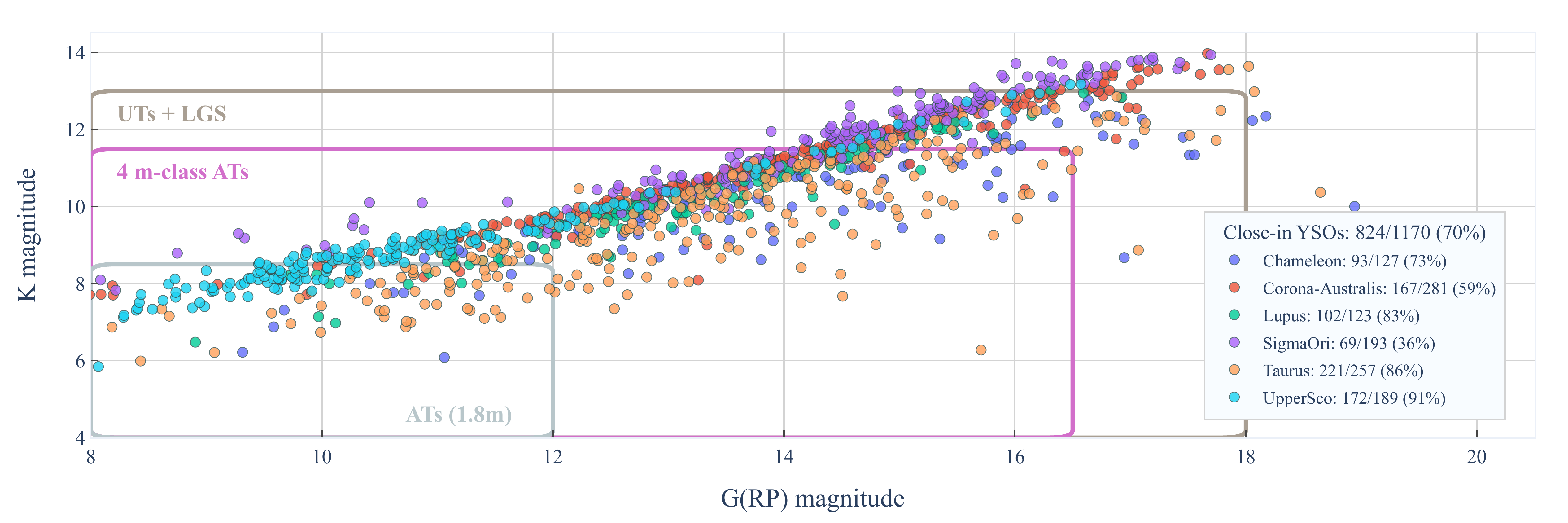} 
     \vspace{-0.5cm}
     \caption{\small Accessible parameter space ($G_{RP}$ vs $K$) for YSOs. The upgrade to 4-m class (violet) bridges the gap between current ATs (gray) and UTs (dark gray), unlocking hundreds of targets for time-domain interferometry.}
     \label{fig:catalog}
     \vspace{-0.5cm} 
 \end{figure*}


The first option proposes deploying fixed or relocatable telescopes on existing AT stations. While exact specifications (count $\ge$ 6, agility, diameter) are outside this work's scope, we assume a baseline of six 4-m telescopes on the AT infrastructure, though our concepts apply generally. This combination unlocks three critical capabilities.

\vspace{-0.3cm}
\subsection{Snapshot imaging \& sensitivity}
Moving from 4 to 6 telescopes increases the number of simultaneous baselines from 6 to 15, and closure phases from 4 to 20. This dense instantaneous $uv$-coverage enables snapshot characterization, allowing us to secure robust geometric constraints in a single epoch without relying on Earth rotation. This effectively ``freezes'' the target's geometry, solving the time-blurring problem \citep{Kraus20}.

Simultaneously, the shift to 4-m apertures bridges the sensitivity gap (reaching within $1.5$ mag of the UTs). This is critical to unlock the vast population of Class I/II stars in nearby star formation region ($\sim820$ sources, Fig. \ref{fig:catalog}).

\vspace{-0.3cm}
\subsection{True operational reactivity}
Unlike the multi-purpose UTs, a dedicated array of 4-m class telescopes offers flexible scheduling that allows for:
\begin{itemize}[noitemsep, topsep=0pt, leftmargin=*]
    \item High-cadence monitoring to track rotational modulation,
    \item Immediate reaction to photometric alerts (within hours),
    \item Tailored observations during specific phases of a burst.
\end{itemize}
\textbf{This transforms the VLTI from a static mapper into a dynamic response engine.}

\vspace{-0.3cm}
\subsection{6D Tomography: From V to N Bands}

To physically interpret the variability, the facility must cover a broad spectral range. Each band probes a distinct physical component:

\begin{itemize}[noitemsep, topsep=0pt, leftmargin=*] \item \textbf{Visible (V): A High-Sensitivity Option.} While core diagnostics lie in the NIR, extending coverage to the visible remains highly desirable. The sheer strength of H$\alpha$ \citep{Alcala17} provides a unique lever to detect the faintest sources and probe high-velocity gas (>100 km/s), offering a powerful complement to infrared tracers.

\item \textbf{Near-IR ($J, H, K$): The Inner Environment.} This primary window reveals the immediate star-disk interaction: the hot dust sublimation rim ($K$), magnetospheric funnel flows \citep[Pa$\beta$, Br$\gamma$,][]{Tessore23, Soulain23}, and ionized winds, effectively linking the accretion shock to the reprocessing dust.

\item \textbf{Mid-IR ($L, M, N$): The Disk Structure.} These bands capture the cool dust emission, chemical composition, dynamical vortices/planet embryos, water snowlines and vertical disk structure \citep{Lopez22}.

\end{itemize}

By coupling this spectral coverage with a \textbf{rapid-response} operational model, the facility enables true \textbf{6D tomography}. We can directly link a burst detected in the optical to its thermal echo in the inner disk (IR), causally connecting the central engine to planet-forming environments.

\vspace{-0.4cm}
\section{The missing piece of the puzzle}
\vspace{-0.2cm}

An interferometer built around six or more 4-m class telescopes represents more than an incremental upgrade; it is a paradigm shift. It will constitute the world's first \textbf{high-sensitivity optical time-domain interferometer}.

In the landscape of the 2030s, the community will possess powerful large-scale spectro-imagers (ALMA, JWST, ELT) and high-cadence imagers (LSST/Rubin, Roman). Yet, a critical gap remains: the ability to \textbf{monitor} the variability processes on the spatial and temporal scales they occur, i.e., from less than 0.1 au on day timescales to a few au on weeks/months timescales. It will act as the \textbf{dynamical counterpart} to the ELT's static resolution and the \textbf{physical interpreter} of large survey photometric alerts.

By unlocking the temporal dimension, this facility enables new science. It allows us to move beyond mapping \textit{where} matter is, to understanding \textit{how} it moves, falls, reacts and evolves to form stars and planets. Finally, by leveraging and upgrading existing ESO infrastructure, this project aligns with the European carbon footprint reduction strategy and minimises the need for new raw materials.

\vspace{-0.4cm}
\begingroup
\tiny
\setlength{\bibsep}{0pt}
\bibliographystyle{aa}
\bibliography{biblio_white_paper}
\endgroup

\end{multicols}

\end{document}